\documentclass[11pt]{article}
\usepackage[utf8]{inputenc}
\usepackage{amsmath}
\usepackage{amsfonts}
\usepackage{amssymb}
\usepackage{amsthm}
\usepackage{graphicx}
\usepackage{wrapfig}

\usepackage{authblk}

\usepackage[left=2cm,right=2cm,top=2cm,bottom=2cm]{geometry}
\usepackage{setspace}
\theoremstyle{definition}

\usepackage{color}

\begin{document}

\title{Multilayer Brain Networks}
\author[1]{Michael Vaiana}
\author[1,*]{Sarah Feldt Muldoon}
\affil[1]{Department of Mathematics and CDSE Program, University at Buffalo, SUNY, Buffalo, NY}
\affil[*]{Corresponding Author: smuldoon@buffalo.edu}
\date{}
\maketitle

\begin{abstract}
The field of neuroscience is facing an unprecedented expanse in the volume and diversity of available data.  Traditionally, network models have provided key insights into the structure and function of the brain.  With the advent of big data in neuroscience, both more sophisticated models capable of characterizing the increasing complexity of the data and novel methods of quantitative analysis are needed.  Recently multilayer networks, a mathematical extension of traditional networks, have gained increasing popularity in neuroscience due to their ability to capture the full information of multi-model, multi-scale, spatiotemporal data sets.  Here, we review multilayer networks and their applications in neuroscience, showing how incorporating the multilayer framework into network neuroscience analysis has uncovered previously hidden features of brain networks.   We specifically highlight the use of multilayer networks to model disease, structure-function relationships, network evolution, and link multi-scale data.  Finally, we close with a discussion of promising new directions of multilayer network neuroscience research and propose a modified definition of multilayer networks designed to unite and clarify the use of the multilayer formalism in describing real-world systems.
\end{abstract}

\section{Introduction}
The human brain is a complex system organized by structural and functional relationships between its elements. Recent experimental advances have resulted in an unprecedented amount of data describing brain structure and function that now allows the brain to be modeled as a network through the measurement of pairwise interactions between its units. This modeling can occur across multiple scales, where the nodes of the network represent the units of the brain, whether they be proteins, neurons, brain regions, or some other measured unit 
\cite{Feldt:2011fh,BassetSporns2017}. Edges of the network represent the strength of connection between two units, and are typically chosen to measure either physical connections (structural networks) or statistical relationships between nodal dynamics (functional networks) 
\cite{Bullmore2009}. The rich theory of networks has been successfully utilized in studying the brain by quantifying network structure though the calculation of descriptive and inferential network statistics which expose otherwise hidden phenomenon.  Measures of centrality, degree distribution, clustering, small-worldness, and more \cite{newman2003structure,boccaletti2006complex,Muldoon:2016cj}  have been used to study disease, task, learning, behavior, and structure 
\cite{SPORNS2004,Bullmore2009,braun2015human,mattar2016brain,BassetSporns2017}.


The success of network theory in uncovering the complex organization of the human brain is not without limits, however, as traditional networks capture only a single mode of interaction between units.  Recording technologies such as functional magnetic resonance imaging (fMRI), magnetoencephalography (MEG), and electroencephalogram (EEG)  capture brain dynamics across time and across multiple frequency bands, and it is important to retain the information of the full frequency spectrum \cite{bassett2006adaptive,liao2013functional,sasai2014frequency, chen2015bold,thompson2015frequency,domenico2016mappingmultiplex} or the full temporal profile \cite{bassett2011dynamicreconfig,Wymbs2012,bassett2013core-periphery,bassett2015learningautonomy} of such recordings.  In addition to measuring functional interactions through fMRI or EEG, structural recording techniques such as diffusion weighted imaging (DWI) measure the presence and strength of physical connections between the various regions of the brain.  The emergence of such increasingly large and multi-modal data sets therefore necessitates a quantitative model that is rich and flexible enough to both describe interactions between multiple scales and modalities and allow for meaningful analysis of the data to provide new and powerful insights into the organization of the brain.  Unfortunately, traditional networks are not equipped to model these multiple interactions across time, frequency, or modality.

In order to meet these new challenges, recent work in network neuroscience has begun to explore the use of multilayer networks to model the multiple complex interactions that traditional networks are not suited to capture.  A multilayer network is a generalization of a traditional network that retains the simplicity of a network yet provides flexibility in modeling multi-modal data.  A mathematical definition is given in section \ref{sec:intro-to-mln}, but intuitively a multilayer network can be thought of as a network of networks, or of a collection of interconnected networks, each of which represents some interaction between its agents.  While the field of multilayer networks is still in its infancy, multilayer networks have begun to be used successfully in diverse topics such as protein interactions \cite{Ou-Yang2014}, ecology \cite{Pilosof2017}, language \cite{martinvcic2016multilayer}, disease \cite{Halu2017}, transportation \cite{gallotti2015multilayer}, and trade \cite{gao2015features} among others.

\begin{figure}[ht!]
\begin{center}
\includegraphics[scale=.5]{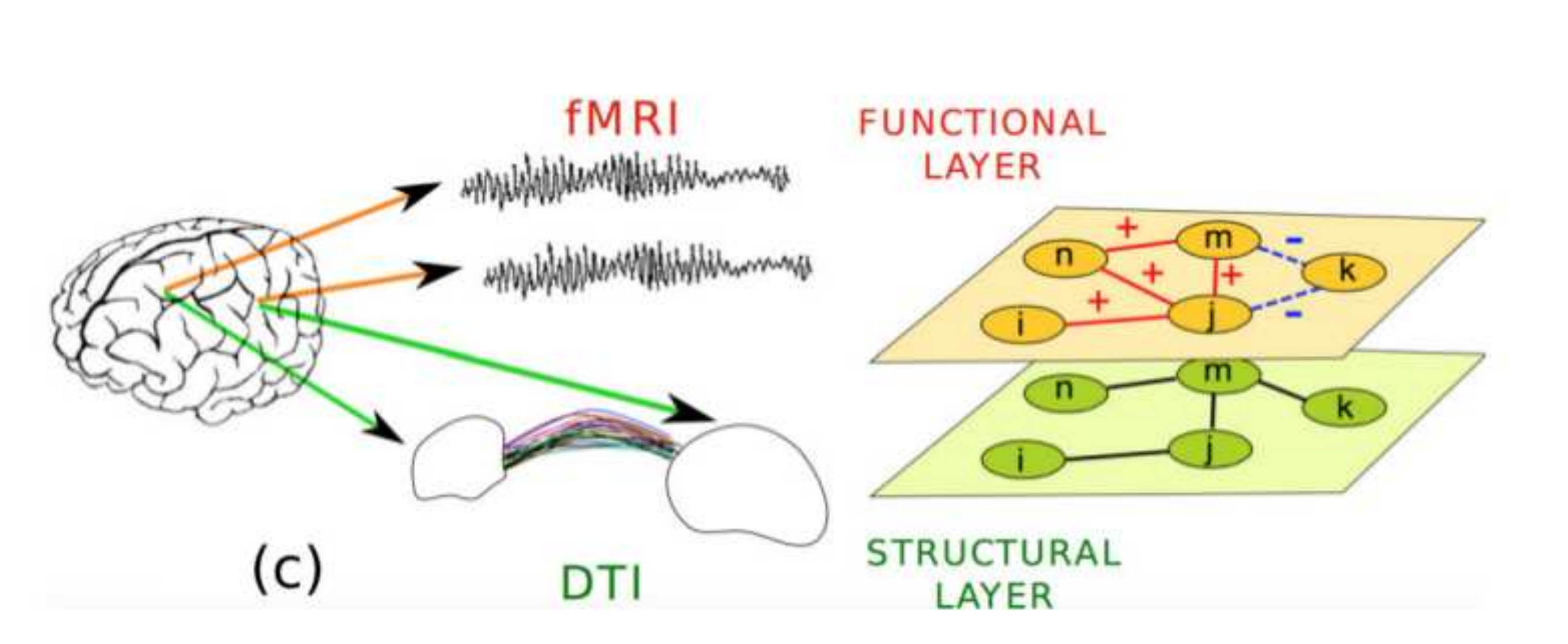}
\caption{A multilayer network created from recording both fMRI and DTI data from a single brain.  Reprinted from \cite{battiston2017multilayer}.}
\label{fig:fmri_mln}
\end{center}
\end{figure}

Recent work in neuroscience has also drawn upon the versatility of the multilayer framework to model the complex relationships in neural data \cite{Muldoon2016}.  For example, given fMRI (functional) and diffusion tensor imaging (DTI; structural) data for a single subject, one can build a multilayer network with two layers: one representing the fMRI network, and the other, the DTI network \cite{battiston2017multilayer}. From the fMRI data, one can build a functional network with the nodes representing brain regions and the edges representing coherence between regional activity.  Given the DTI data, one can build a structural network by again parcellating the brain into regions and then measuring the strength of physical connections between these regions.  Finally the multilayer network is formed by considering each of these networks as a \emph{layer} of the multilayer network and adding edges from a brain region in the fMRI layer to itself in the DTI layer (see Fig. \ref{fig:fmri_mln}).

In this review, we first introduce the general concept of multilayer networks.  We next describe applications of the multilayer framework to problems in neuroscience, followed by a discussion of currently available network statistics for use in the analysis of such data.  Finally, we offer insight into future directions of multilayer networks and their use in neuroscience, and present a modified definition of a multilayer network meant to unify research in this developing field.

\section{General Introduction to Multilayer Networks}
\label{sec:intro-to-mln}
\subsection{Mathematical Definition}
\label{sec:sub:def-and-example}
In the following, we provide a general introduction and mathematical definition of multilayer networks.  For a more detailed description of multilayer networks, please see the comprehensive review by Kivela et al. \cite{Kivela2014}.

In order to describe a multilayer network, we first review the definition of a traditional network.  A traditional network, $N$ can be formalized as a tuple $N=(V,E,w)$ where $V$ is a set of nodes, $E = V\times V$ is a set of edges, and $w:E\to R$ is a function on the edges called a weight function. It is clear from this definition that a network can be represented as a matrix, $A$, with $A_{ij} = w(v_{i},v_{j}).$    This matrix is called the \emph{adjacency matrix} of the network.   One may specialize to a certain class of networks by imposing constraints on the weight function $w$.  For example, if one assumes $w$ is symmetric, i.e., $w(v_1,v_2) = w(v_2,v_1)$ then the network is said to be undirected. When edge weights are restricted to binary numbers (i.e., $R=\{0,1\}$) the network is said to be unweighted.  Because the full information about the network is contained in the adjacency matrix, we often use the formal definition of the network and its adjacency matrix interchangeably.

A \emph{multilayer network} builds upon the traditional network with the addition of an extra labeling function, $l:V\to L$ from the nodes to a set of labels, $L,$ i.e. $\mathcal{M} = (V,E,w,l).$  The labeling set is assumed to take the form $L = L_1\times L_2 \times \ldots \times L_d$, where each set $L_i$ is called a dimension, or aspect, of the multilayer network, and the network is said to be $d$-dimensional.  Mathematically, the only difference between a network and a multilayer network is that in a traditional network, each node is assigned a label $v\in V$, while in a multilayer network, each node is instead assigned a vector label $l(v) = (v_1, v_2, \ldots, v_d)$. The true power of multilayer networks is the ability to select subsets of nodes, which we call layers, based on their labels.

Thus, in a multilayer formalism, one can essentially build multiple layers of traditional networks, where each layer describes some feature (aspect/dimension) of the data.  Intra-layer edges serve the same purpose as in a traditional network, describing network connections which represent a single feature of the data, while inter-layer edges provide connections between features.  This allows a multilayer network to capture several modes of interaction between its nodes.


We must also clarify our vocabulary when we discuss network nodes in a multilayer framework.  For notational convenience we identify each node $v\in V$ with its label \[v \sim \mathbf{v} = l(v)= (v_1, v_2, \ldots, v_d).\]  However, this can lead to confusion between nodes in different layers of the network.  For clarity, we will therefore make a distinction between network nodes and vertices in our multilayer setting.  The nodes should be thought of as the unique objects of interest in the network, for example brain regions or neurons, and the vertices should be thought of as the abstract underlying vertex set of the full multilayer network.  A node can be present in all layers of the network, but does not have to be (e.g., if nodes represent neurons and layers represent time, new neurons could be born over time and old neurons could die).

\begin{figure}[ht!]
\begin{center}
\includegraphics{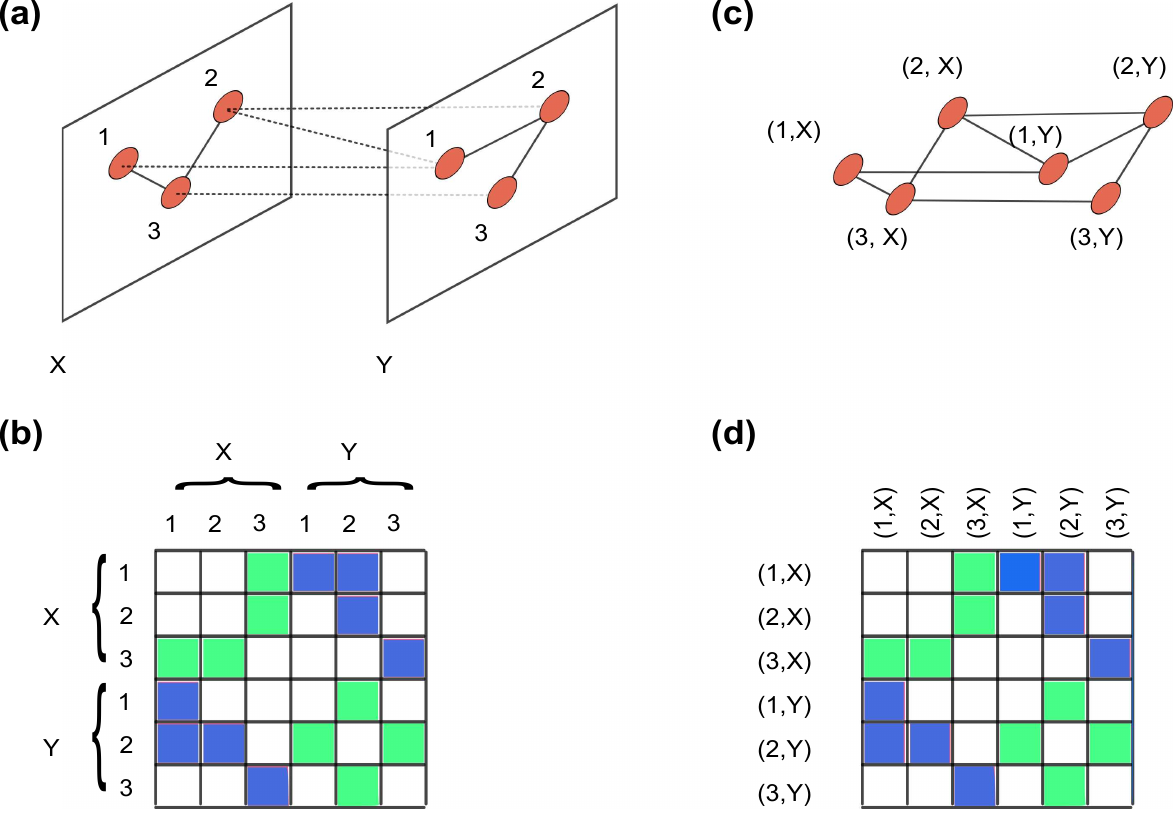}
\caption{Four representations of the same multilayer network. The adjacency matrices are shown at the bottom with intra-layer and inter-layer edges colored green and blue respectively and white representing no edge present (color online). \textbf{(a)} The multilayer network visualized with an emphasis on the partition of the network into layers. Intra-layer edges are solid and inter-layer edges are dashed. \textbf{(b)}.  The adjacency matrix labeled with an emphasis on the partition of the network into layers.   \textbf{(c)} The multilayer network shown as a network whose nodes are vector labeled.  \textbf{(d)}  The vector labeled adjacency matrix.  }
\label{fig:mln_with_adjmat}
\end{center}
\end{figure}

Mathematically, we will denote $L_1 = N$ as the set of node identities representing physical objects.  We will therefore call the elements $i \in N$ the \emph{nodes} of a network and $v\in V$ the \emph{vertices} of the network to distinguish between them. This distinction is visually depicted in Fig. \ref{fig:mln_with_adjmat}, where panels (a-b) highlight the nodes of the network partitioned into layers, and panels (c-d) showcase the vertex set of the underlying network.

Multilayer networks can also be represented by a matrix in the same way as a traditional network, although in the multilayer framework, we refer to this matrix as the supra-adjacency matrix.  For every pair of vertices in the network $v_i, v_j$ we have a weight, $w(v_i, v_j)$, that describes the edge between $v_i$ and $v_j.$  If $A_{ij} = w(v_i, v_j)$, then the pair $(A, l)$ defines the supra-adjacency matrix, where $l$ is the labeling function of the multilayer network.  We often suppress the function $l$ but maintain its information by labeling the $i$th row of $A$ by the vector $l(v_i)$ (see Fig. \ref{fig:mln_with_adjmat}).

The supra-adjacency matrix is a popular data structure for multilayer networks.  It retains the full information of the network and the layers of the network appear in the supra-adjacency matrix as blocks.  The diagonal blocks consist of intra-layer edges, and the off diagonal blocks consist of inter-layer edges.  The supra-adjacency matrix has the nice property that, up to row labels, it is identical to the adjacency matrix of the underlying network.  Many methods of analysis on multilayer networks reduce to careful manipulation of this matrix.  However, unlike in a traditional network the rows/columns of the supra-adjacency matrix do not simply represent network nodes -- they represent the full set of vertices.  Therefore, it is essential to equate each row/column of the adjacency matrix with the full vertex label as in Fig. \ref{fig:mln_with_adjmat}(c-d).  (Note that multilayer networks can also be represented using tensor notation \cite{domenico2013mathematical}, but we will not use this notation here.)

\subsection{Example: Temporal Networks}
\label{ex:temp_net2}
One of the most common applications of the multilayer formalism to neuroscience data thus far has been through defining temporal networks based on time series data collected from multiple brain regions or electrodes.  A temporal network is an example of a multilayer network with two dimensions: nodes and time.  Each layer depicts the state of the network at a specific point in time, and the layers are sequentially linked, representing the flow of time.  As an example, consider a data set that consists of the activity of 50 brain regions measured over 10 time windows. Then there are 500 data points: one for each brain region in each time window. We can build a multilayer network such that $V$ consists of the 500 data points and $l:V\to N\times T$ with $N = \{1,2,\ldots, 50\}$ and $T = \{1,2,\ldots, 10\}.$  Then $\mathbf{v} = (n,t)$ identifies a data point with a brain region at a specific time.

\begin{figure}[ht!]
\begin{center}
\includegraphics{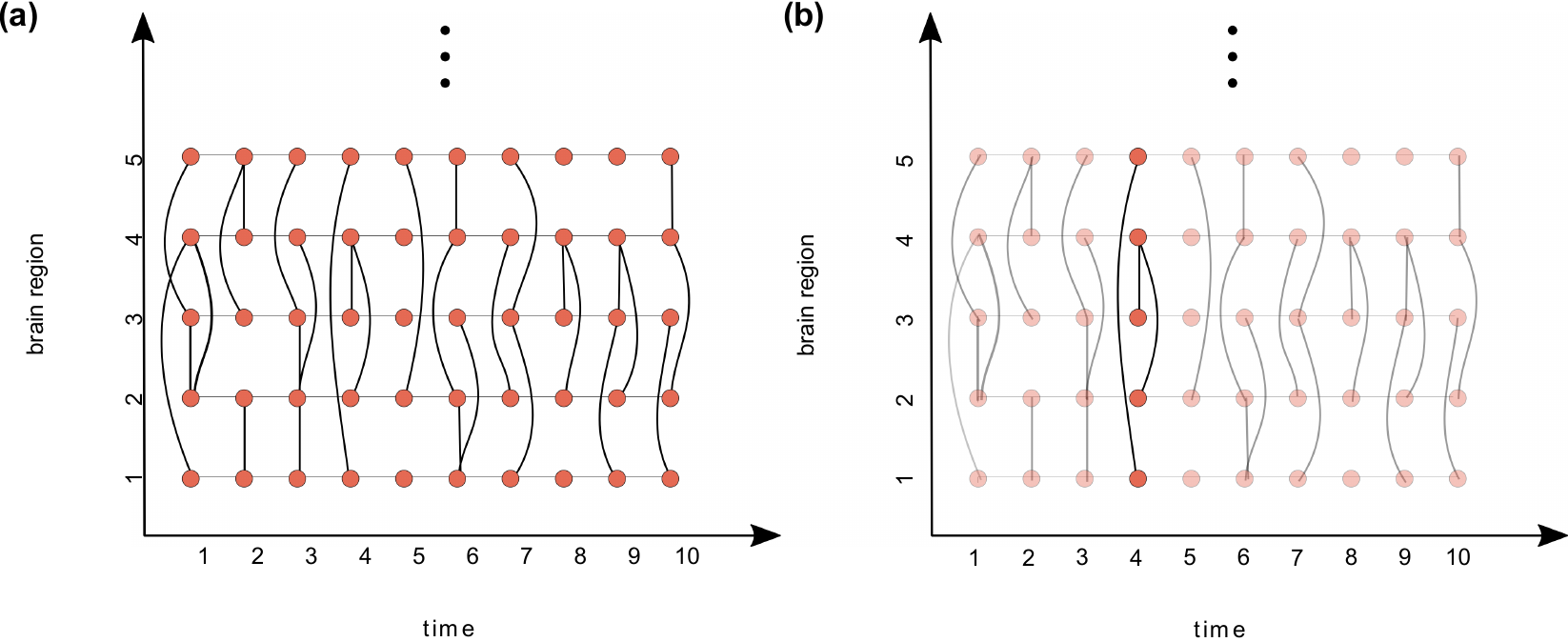}
\caption{A visualization of a temporal network. Each layer of the network corresponds to a time window.  intra-layer edges run vertically and inter-layer edges run horizontally. \textbf{(a)}. A temporal network with 10 time layers. \textbf{(b)}. The same network as in (a) where the layer of the network corresponding to time window 4 has been highlighted.}
\label{fig:layers}
\end{center}
\end{figure}

For each time window, we can define the intra-layer edge between region $i$ and region $j$ to reflect an empirical measure of similarity between the activity of the two regions in that time window.  Inter-layer edges between time windows are added only between a node and itself in an adjacent time window (i.e., there is an edge between $(m,s)$ and $(n,t)$ if and only if $m = n$ and $s= t \pm 1$). The weight of each inter-layer edge is generally set to some constant, $\omega$, which then becomes a parameter of the network.  In Fig. \ref{fig:layers} the temporal network is visualized as a two dimensional network with intra-layer edges running vertically and inter-layer edges running horizontally. For compactness only the first 5 out of 50 brain regions are shown.

\subsection{Utility of Multilayer Networks}
As we have stated, a multilayer network is simply an extension of the traditional network in which network nodes are labeled by vectors.  In light of this, it is natural to ask what is the utility of a multilayer network? Why not just model each aspect of the data as an independent network?  One power of a multilayer network is, in fact, mostly conceptual and organizational.  The vector labels give identifications of meaningful subsets of the full data set.  The analytic power of a multilayer network comes from comparing statistics across these subsets.  Until recently when one was presented with network data that was inherently multidimensional, a common approach was to aggregate this information into a single layer network, for example by averaging.  With the advent of multilayer networks and network statistics designed to operate across layers, the full multidimensional information can be retained, allowing for a richer analysis that could not be obtained from the aggregated single layer network. For example, in \cite{domenico2016mappingmultiplex}, the authors show that a multilayer representation of a human brain network gives higher classification accuracy between healthy and schizophrenic patients than the single layer and aggregate counterpart networks.

Considering the inverse problem further illuminates the power of a multilayer network.  Starting with the data of a network and the goal of making comparisons across meaningful subsets of the network, it is first necessary to find those meaningful subsets, a problem without a clear or easy solution.  A multilayer network, however, comes with distinguished subsets from which one can easily make comparisons. These subsets have meaning which reflects the experiments or data from which they were derived.  Data often naturally comes partitioned, for example by sampling multiple frequencies, times, subjects, or scales \cite{Muldoon2016,BassetSporns2017,DeDomenico2017}.  The multilayer network is a natural model for these data sets.

Perhaps the most compelling reason for utilizing the multilayer framework is its ability to provide a coherent and linked representation of network attributes across aspects.  Consider the simple example of a temporal network.  It is reasonable to believe that in time series data, one time point (window of data) is not independent from the previous point.  The inter-layer edge weight, $\omega$, allows one to control the strength of interaction between layers, i.e., how much influence one layer has on the next.  In the context of a temporal network, this also means that one can coherently track network properties throughout time.  This is especially apparent when performing dynamic community detection, which we discuss in detail in section \ref{evolution}.  One is able to define and track network communities throughout time, allowing for a much richer description of the evolution of network structure: communities can be born, communities can die, and nodes can change their community affiliation over the course of time.

\section{Multilayer Networks in Neuroscience}
\label{sec:use_of_mln}
How does the brain change with learning?  What are the neurobiological markers of diseases such as autism, Alzheimer's disorder (AD), epilepsy, or schizophrenia, and can we provide better interventions or diagnoses? How are the neurons in our brain connected to each other, and how does this relate to brain function?  These are only a handful of broad questions of interest to the neuroscientist. As we have already emphasized, the investigation of these questions naturally spans multiple spatial and temporal scales, results in the use of multiple modalities to record data, and often involves comparison across multiple subjects or between subject groups. A multilayer network is a suitable tool for reconciling these data in a coherent and consistent manner.  Its nodes (e.g. brain regions) can be coherently linked across many modes of analysis resulting in a single structure which encapsulates the full information of the data.  As we will see shortly, new insights into the above questions have been found through the use of multilayer networks.

\subsection{Disease}
Networks of the brain have long been used to study neurological disease \cite{braun2015human,Bassett:2009ct,Fornito:2015dq}, but new methods using multilayer analysis of data from healthy and diseased patients have recently been employed to study a variety of disorders. The comparison of multilayer network measures between control (healthy) groups and disease groups can provide powerful insights into multiple disorders.

\begin{figure}[ht!]
\begin{center}
\includegraphics[width=0.9\textwidth]{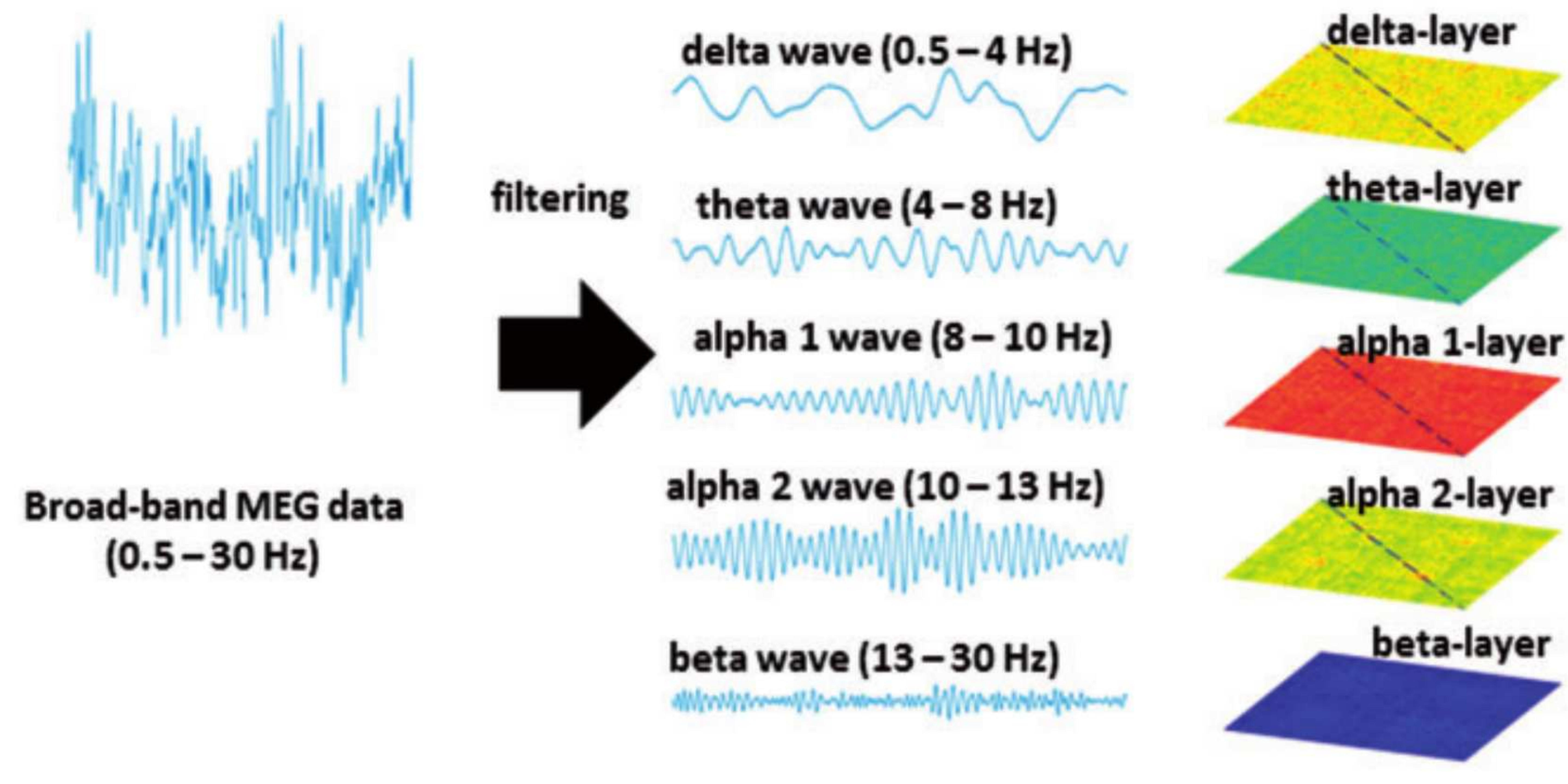}
\caption{Constructing a multilayer frequency network from an MEG signal. Reprinted from \cite{yu2017selective}.}
\label{fig:yu_2017}
\end{center}
\end{figure}

One paradigm for creating a multilayer network from fMRI, MEG, or EEG data, is to first decompose the signal into several frequency bands for each brain region and then measure functional similarity between brain regions in each of these frequency layers (see Fig. \ref{fig:yu_2017}).  Inter-layer edges can be added by different means, such as by coupling all brain regions to themselves across layers \cite{domenico2016mappingmultiplex,yu2017selective} or by measuring similarity between signals across layers \cite{brookes2016multi}.  This is in direct contrast to traditional approaches which either first select a frequency of interest and then construct a single layer, frequency specific network, or consider all layers corresponding to distinct frequencies but then aggregate them into a single network by summing edges across layers. In the case of fMRI, it has been indicated that important information may be contained in typically excluded frequency bands \cite{bassett2006adaptive,chen2015bold,liao2013functional,thompson2015frequency}, motivating the need to retain the full frequency information during analysis.

Single layer networks have been been successfully employed in the study of AD \cite{Supekar2008,tijms2013alzheimer} and schizophrenia \cite{van2014brain,lynall2010functional,lee2004social,calhoun2009functional} and recently multilayer frequency networks have been applied to the study of these diseases \cite{yu2017selective,domenico2016mappingmultiplex,brookes2016multi}. The \emph{centrality} of a node in a network is a measure of its importance within the network, and measures of centrality have been leveraged to uncover interesting brain organization.  Many measures of centrality, including PageRank \cite{page1999pagerank} have been extended to the multilayer setting \cite{sola2013eigenvector,battiston2014structural,domenico2015ranking,iacovacci2016extracting}. In \cite{domenico2016mappingmultiplex}, multilayer PageRank centrality \cite{domenico2015ranking} is used as an input to a classifier which is able to discriminate between healthy and schizophrenic individuals significantly better than the single layer counterparts.  In addition, multilayer \emph{hubs} (highly central nodes) were used to identify regions distinctive to schizophrenic brains.  In \cite{yu2017selective}, frequency specific and multilayer frequency networks were constructed from MEG data to investigate AD.  Using a measure of degree centralility \cite{battiston2014structural}, the multilayer networks in patients with AD were characterized by loss of nodal centrality compared to the healthy group where as no such difference was found in the frequency specific networks.  This was the first MEG study that identified selectively vulnerable hubs in AD, and importantly, these hubs are consistent with hubs reported in previous studies using DTI, MRI and fMRI data.  This suggest that a multilayer approach may be essential in future analysis of MEG data.

\subsection{Structure}
The structural organization of the brain of even the most elementary species is immensely complex with an enormous number of connections.  To date, \emph{C.elegans} is the only species for which we have a complete structural map, consisting of 302 neurons and approximately 7000 connections \cite{white1986structure}.

\begin{wrapfigure}{r}{0.4\textwidth}
\begin{center}
\includegraphics[width=0.38\textwidth]{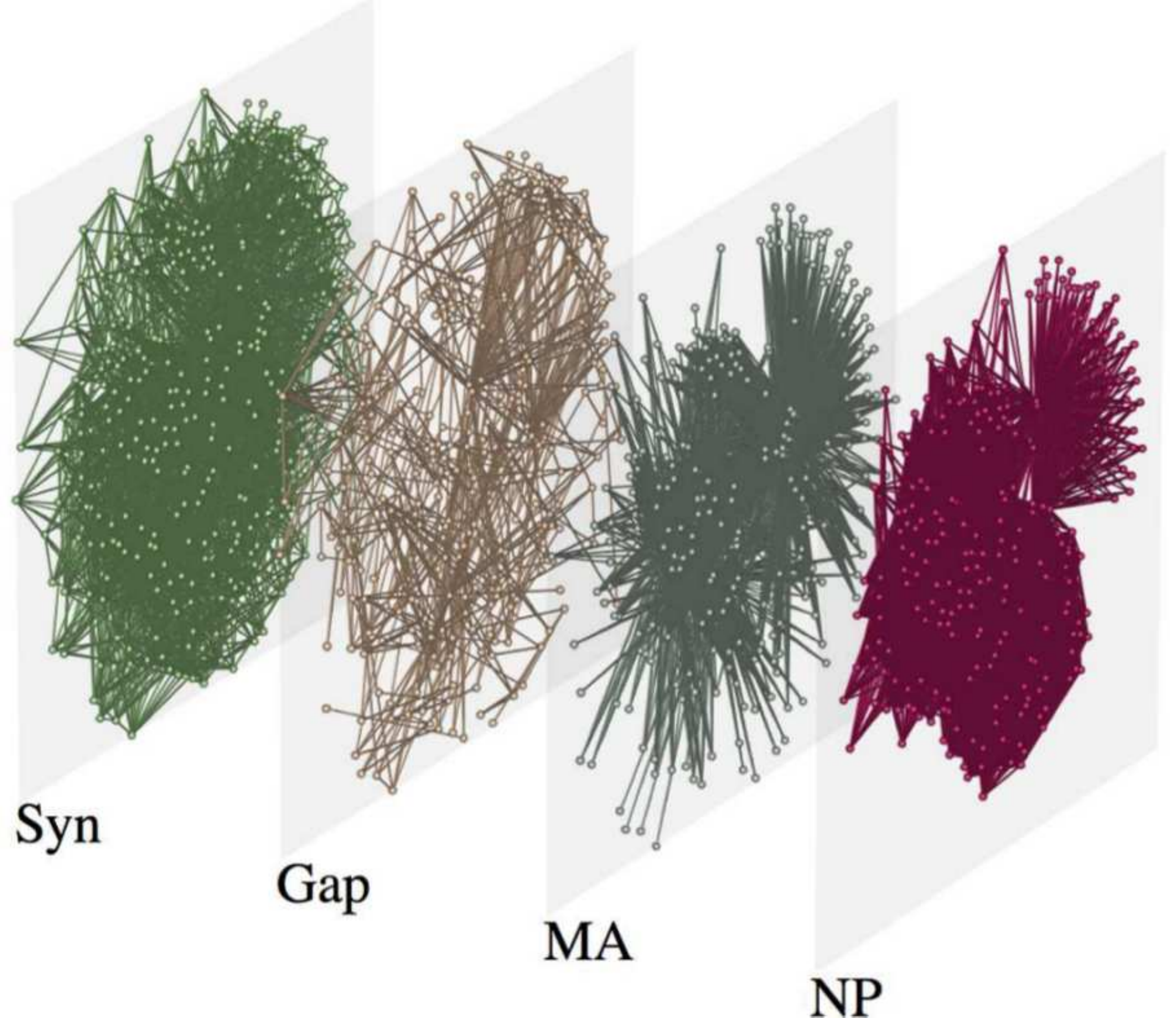}
\caption{The 4 layers (synaptic, gap junction, monoamine, neuropeptide) of the multilayer neural communication network of \emph{C. elegans}.  Reprinted from \cite{Bentley2017}}
\label{fig:bentley}
\end{center}
\end{wrapfigure}

As with functional data, structural data is inherently multiplex, with structure occurring at different scales and in different modes.  For example, structural connections are measured at the micro-scale cellular level between neurons (as with \emph{C. elegans}) and are also measured at the meso-scale level by estimating the number of \emph{white matter tracts} connecting different brain regions.  Even for a fixed level of measurement, for example the cellular level, there are multiple modes of structural connections between neurons, i.e. synaptic and extra-synaptic connections. A partial multilayer connectome in which each layer represents a mode of interaction (synaptic, gap junction, monoamine, and neuropeptide) between the neurons of \emph{C. elegans} was constructed in \cite{Bentley2017} (see Fig. \ref{fig:bentley}).  Multilayer measures of network topology \cite{battiston2014structural}, including clustering, assortivity, path length, modularity and motif analysis were employed to probe the structure of this network. In particular, it was shown that synaptic and extra-synaptic connection networks exhibit distinct network topologies and simultaneously share certain network multilayer motifs and hubs indicating that these networks may work both independently and in unison.

In addition to understanding the structure of the nervous system it is important to understand the relationship between structure and function. A network motif is a subgraph of the network which occurs more frequently than expected by chance.  In \cite{battiston2017multilayer} the authors construct a multilayer network consisting of a structural and functional network (see Fig \ref{fig:fmri_mln}) and analyze the motif structure of this network.  Their results indicate that the functional brain network is constrained by its structure and that structural connections may be necessary for positive functional correlations.

\subsection{Network Evolution}
\label{evolution}
It is often necessary to ask how brain networks evolve over time, as many brain functions such as learning are by definition temporal processes.  For example, a subject can progress from naive to advanced stages of learning over the course of days or weeks.  Characterizing the mechanisms, signatures, and predictors of learning is a broad and fundamental goal of neuroscience.  As previously discussed, a temporal network is a natural model for capturing brain dynamics through time.  Like the other uses of multilayer networks previously mentioned, a temporal network has the advantage of retaining the full information of the data without the need to aggregate connections into a single network.

Multiple studies have used temporal networks to investigate the \emph{dynamic community structure} of brain networks to reveal otherwise hidden phenomenon in human brain dynamics during learning \cite{bassett2011dynamicreconfig, Wymbs2012, bassett2013core-periphery, bassett2015learningautonomy, Braun2015,Telseford2016, Brovelli2017,telesford2017cohesive}.  These studies build functional temporal networks using fMRI data by parcellating the brain into regions and calculating functional similarity for intra-layer edges and using temporal inter-layer coupling as described in section \ref{ex:temp_net2}.  Communities (groups of nodes with strong internal connections and weak external connections) have played an important role in network neuroscience (see \cite{Sporns2016} for a review of communities in brain networks) and also play a crucial role in exposing underlying phenomenon in temporal networks.

Detecting communities in networks is a difficult problem with a long history of research (see \cite{Porter2009communities} for a review).  Several methods of community detection have been extended to multilayer networks \cite{Mucha2010,Gauvin2014,domenico2015identifying,matias2016statistical,Paul2016,boutemine2017mining,zhang2017modularity}.  A popular method of community detection that is commonly used in neuroscience is the method of maximum modularity \cite{Newman2006}, which uses a Louvain like heuristic \cite{Blondel2008, Jutla2011} to maximize a quality function. The method of maximum modularity, or modularity for short, has known problems \cite{Good2010} yet can be made robust \cite{bassett2013robust} and performs well on benchmark graphs \cite{Yang2016}. The heuristic for modularity maximization  is  fast  and the use of multilayer modularity has provided good results in practice \cite{bassett2011dynamicreconfig, bassett2013core-periphery, bassett2015learningautonomy, Braun2015,Telseford2016}.

 In the multilayer setting \cite{Lambiotte2008,Mucha2010}, the modularity function takes the form
 \[Q = \sum_{ijrs} [(A_{ijs} - \gamma_s P_{ijs})\delta(s,r) + \omega_{jsr}]\delta(c_{ir}, c_{is}).\]
 Here, $A_{ijs}$ is the adjacency matrix of layer $s$, and $P_{ijs}$ is the associated null matrix. The term $\gamma_s$ is a resolution parameter and $\omega_{jsr}$ is the link between node $j$ in layer $r$ and node $j$ in layer $s.$  In functional dynamic brain networks the inter-layer edges $\omega_{jrs}$ are not typically measured empirically and are set as
 \[\omega_{jrs}=
 \begin{cases}
 \omega \quad s = r \pm 1\\
 0 \quad \textrm{otherwise}.
 \end{cases}
 \]
 Thus, nodes are only connected to themselves in adjacent time windows and all inter-layer edges are set to the same value $\omega.$  In this way $\omega$ becomes another parameter of the modularity function.  The multilayer modularity function therefore has three parameters: the resolution parameter $\gamma,$ the inter-layer coupling parameter $\omega,$ and the null network $P_{ijs}.$  We mention that other methods of multilayer community detection do not rely on null models \cite{domenico2015identifying} or user-tuned parameters \cite{boutemine2017mining} and that in practice the method of community detection chosen should be deliberate, and the user should be aware of the benefits and drawbacks of each method.

\begin{figure}[ht!]
\begin{center}
\includegraphics[width=0.9\textwidth]{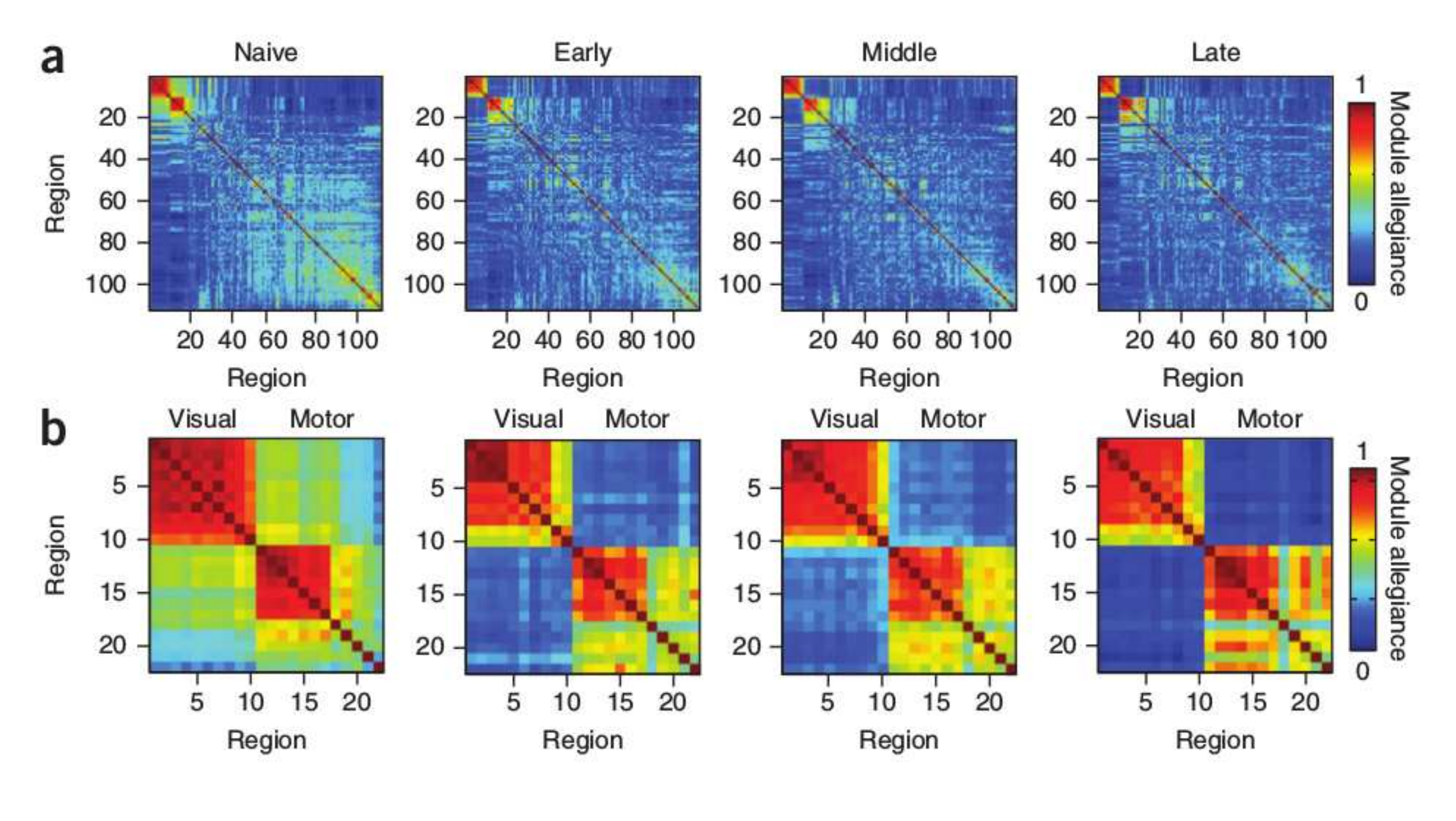}
\caption{The module allegiance matrix for 4 different scanning sessions. Subjects were scanned while performing a motor task over 4 different sessions (naive, early, middle, and late).  In between scanning sessions subjects practiced the motor tasks at home. \textbf{a.} The full module allegiance matrix calculated across all subjects.  \textbf{b.} A zoomed in view of the module allegiance matrix corresponding to only the visual and motor systems. Reprinted from \ref{fig:basset2015learningautonomy}. }
\label{fig:basset2015learningautonomy}
\end{center}
\end{figure}
Because multilayer community structure defines the evolution of communities throughout time, novel network statistics have been developed to exploit this new network feature. For example, one can ask questions about a node's community affiliation over time.  Flexibility, $f_i$, is defined as the number of times a node node changes its community assignment divided by the total possible number of such changes \cite{bassett2011dynamicreconfig}.  The flexibility of the network is then defined as the average flexibility of all its nodes: $F = \frac{1}{N}\sum_i f_i$. Flexibility has been used to measure a core-periphery structure in brain networks, where nodes that have significantly high flexibility are considered periphery nodes, whereas nodes with significantly low flexibility are considered core nodes \cite{bassett2013core-periphery}.  Flexibility and core-periphery structure have then been linked to the adaptive architecture of the brain in task performance and learning \cite{bassett2011dynamicreconfig,bassett2013core-periphery,Braun2015,Telseford2016}.  For example in \cite{bassett2013core-periphery} the authors show that individuals with stronger core-periphery structure learn a simple motor task better than individuals with weaker core-periphery structure.
\begin{wrapfigure}{r}{0.4\textwidth}
\begin{center}
\includegraphics[width=0.38\textwidth]{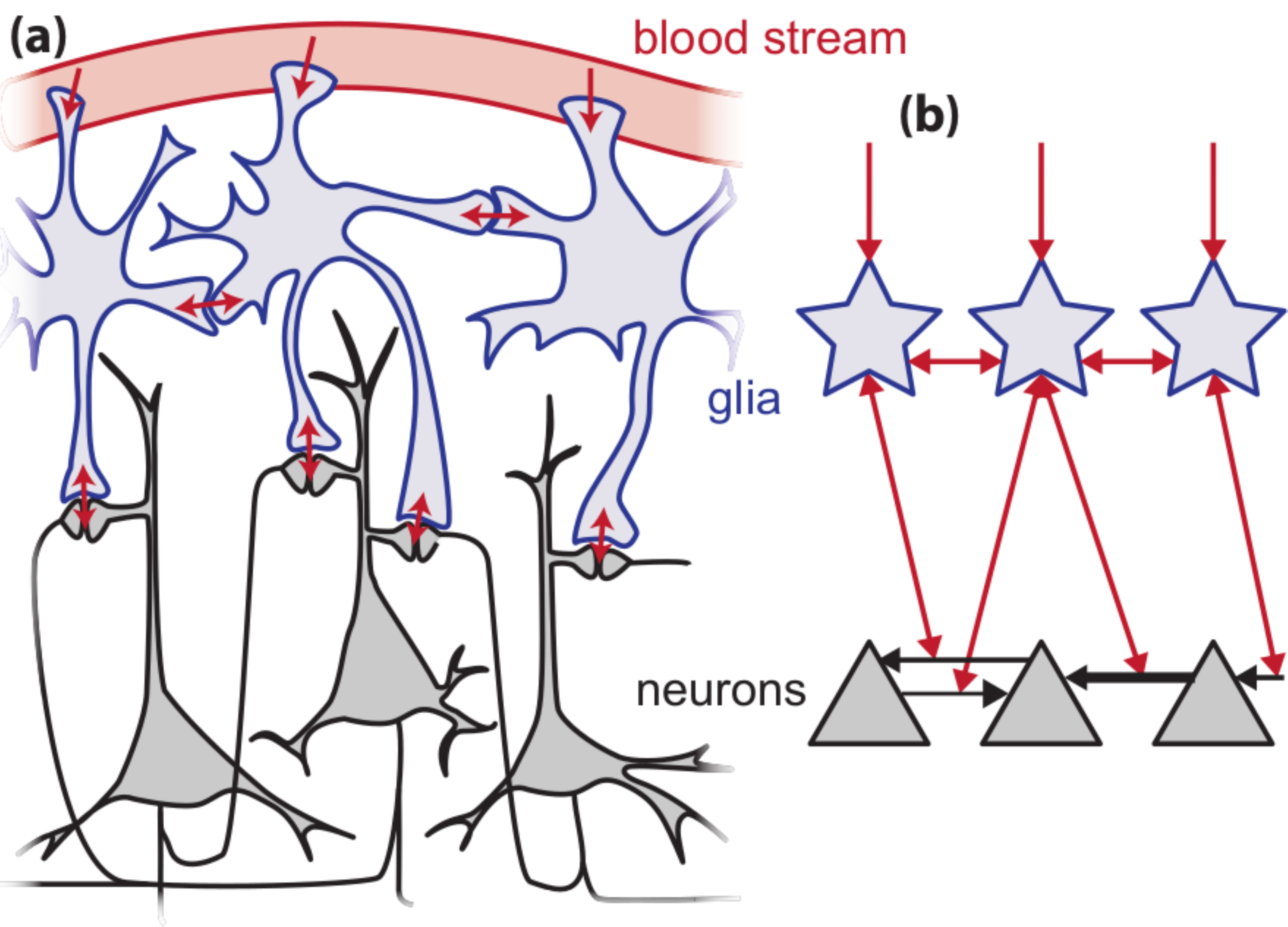}
\caption{\textbf{a.} A cartoon illustrating how glia serve to distribute resources neural synapses. \textbf{b.} A simplified graph representing the two layer glia-neuron network model.  Reprinted from \cite{virkar2016feedback}}.
\label{fig:glia}
\end{center}
\end{wrapfigure}

Another novel measure of network structure resulting from employing multilayer community detection is the module allegiance matrix, $\mathbf{P}$.   Let $T_{ij}$ be a matrix whose $i,j$ entry is the number of times that node $i$ and $j$ were assigned to the same community over all the layers of a multilayer network.
Then $\mathbf{P}_{ij}=\frac{T_{ij}}{C}$, where $C$ is the total number of layers.
Thus $\mathbf{P}_{ij}$ is the probability that $i$ and $j$ are in the same community across the layers. This can be defined even more broadly. Given $N$ communities labels per node, then define $\mathbf{P} = \frac{\mathbf{T}}{N}.$  For example, in \cite{bassett2015learningautonomy} the authors measure the fMRI signal for a number of subjects over a number of scanning trials per subject.  For each scanning session a dynamic functional brain network is formed and $T_{ij}$ is the number of times brain region $i$ and $j$ are assigned the same community over all layers, scans, and subjects.
Analysis of the structure of the module allegiance matrix shows that as subjects learn a task, the motor and visual system decouple.  Further, the non-motor and non-visual system becomes less integrated through training (see Fig. \ref{fig:basset2015learningautonomy}). 
 This supports the intuitively plausible idea that cognitive control is critical during skill acquisition and becomes less so as skills become automatic.  Importantly, the decoupling of the motor and visual system was not detectable through the fMRI signal proving the module allegiance matrix, whose input includes a temporal network for each subject and each scan, is essential in detecting this dynamic configuration of the brain during learning.

\subsection{Linking scales}
Learning is not only a temporal process but is also a multiscale process.  Synaptic plasticity, the microscale process through which neurons change their synaptic connections, is thought to play a key role in learning.  In \cite{virkar2016feedback} the authors build a two layer network consisting of neurons and \emph{glia}, cells responsible for distributing metabolic resources to neural synapses (see Fig. \ref{fig:glia}).
 Using this multilayer model, the authors show that the glia resource distribution can serve as a simple, realistic mechanism for preserving the stability of learning neural systems.

\section{Multilayer Network Statistics}
In the previous section, we described examples of how multilayer network analysis has provided novel insights into neuroscience data.  These studies both relied on the application of traditional network statistics that have been extended to the multilayer framework (PageRank centrality \cite{domenico2015ranking}, degree centrality \cite{battiston2014structural}, modularity\cite{Mucha2010}), as well as led to the development of novel network statistics designed specifically for multilayer applications (flexiblitity \cite{bassett2011dynamicreconfig}, core-periphery \cite{bassett2011dynamicreconfig}, integration and recruitment \cite{bassett2015learningautonomy}).

While they have not necessarily been utilized in the context of neuroscience, many more measures for multilayer networks exist.  Multilayer networks have been studied in the context of
community structure \cite{Lambiotte2008,Mucha2010,domenico2015identifying,Bazzi2016,Bazzi2017,bassett2013robust,bassett2011dynamicreconfig,bassett2013core-periphery,bassett2015learningautonomy,taylor2016enhanced,Sarzynska2014,Gauvin2014,zhang2017modularity,matias2016statistical,iacovacci2015mesoscopic,Papadopoulos2016},
centrality
\cite{domenico2015ranking,sola2013eigenvector,domenico2016mappingmultiplex, iacovacci2016extracting},
percolation and dynamics
\cite{DeDomenico2016,arenas2016nonlinear,khambhati2017modeling,majhi2016chimera,makarov2016emergence,Boccaletti2014,Maksimenko2016,cellai2013percolation}, motifs
\cite{battiston2017multilayer},
and (broadly) structure
\cite{battiston2014structural,nicosia2017collective,arruda2016degree,rombach2017core,danziger2016effect,Kivelae2015,domenico2015structural}.  As the multilayer framework is increasingly incorporated into studies involving network neuroscience, research will benefit from incorporating more of these measures into the analysis of brain networks.

\section{Future Directions}
Network science is poised to be a primary tool of quantitative analysis of brain networks \cite{Muldoon:2014ci, BassetSporns2017, DeDomenico2017,Medaglia:2015jma} and the future of network science will be driven by the goal of understanding quantitative and qualitative properties of multilayer networks and the dynamics that take place on them \cite{Muldoon2016}.  Here, we highlight some specific areas of active research that will benefit both general networks researchers and neuroscientists alike.

\subsection{Novel modeling paradigms}
In section \ref{sec:use_of_mln} we detailed some of the common paradigms for multilayer networks in neuroscience, including networks built with frequency layers and temporal networks built with time layers.  Each paradigm has provided powerful insights and was explicitly shown to reveal more information than its single layer counterpart.  To date, we are unaware of any study that has utilized a three dimensional network with labeling set, $N\times F \times T$ (nodes, frequency, and time), which has the potential to exploit the power of both the frequency and temporal networks simultaneously. Using the definition and ideas in section \ref{sec:sub:unifying_definition} to conceptually frame the analysis of this three dimensional network, this model has the potential to provide new and exciting results.  We note that it is possible to build such a network by simply combining the methodologies employed to build each the frequency and temporal networks so constructing such a network poses no technical challenges.

\subsection{Null Models}
One of the powers of network theory is its ability to compare a measure against the same measure on a random network.  Traditional network theory has developed several useful random network models each based on different assumptions.  For example the configuration model \cite{luczak1989sparse} is a randomized version of a given network that preserves the degree sequence.  In cases where the degree distribution is an important feature of a network, one can use this model to compare a measure for a given network to that of a random network exhibiting the same degree distribution.  Other examples of poular null models include the Erdos-Reni \cite{erdos1960evolution} random graph and the stochastic block model\cite{decelle2011asymptotic}.

In order to discover meaningful information in multilayer networks it is also important to compare a given measure to that of a null model for the network just as is the case for traditional networks.  Some steps in this direction have been taken \cite{bassett2013robust,Sarzynska2014,Paul2016,Bazzi2016,Bazzi2017} but more research is needed.  It will be particularly important to develop null models that are specifically designed to represent the features of neural data, perhaps by considering spatial and geometrical constraints naturally imposed on the brain.

\subsection{Inter-layer Edges}
\label{sec:sub:inter_edges}
The role of inter-layer edges in a multilayer network is simple but critical.  These edges serve to tie together several disparate networks, usually through means of an identity link, which connects a node to itself in a different layer.  While it is possible to experimentally determine the strength of inter-layer edges, as opposed to considering them simply as a model parameter, few studies to date have taken this approach \cite{brookes2016multi}.  Future work will be needed to understand the affects of inter-layer coupling on the structure and dynamics of the resulting network.  While some work has been performed in the context of community detection that indicates that it is possible to set the value of inter-layer edges as a parameter in a principled way \cite{bassett2013robust,Bazzi2016}, the parameter is not well understood.  Even less is understood about the difference between experimentally measuring inter-layer edges (when possible) or considering inter-layer edges to be a tunable model parameter.  Even in the case that inter-layer edges are measured experimentally, it is still possible to weight their importance relative to intra-layer edges by a global parameter.  Future work is needed in order to understand the effect these edges play in different network measures including, for example, community structure and centrality.

\subsection{Unifying Definition}
\label{sec:sub:unifying_definition}
Finally, we believe a direction of future research in multilayer networks should consist of a general and unifying definition of the object of study.  In section \ref{sec:sub:def-and-example}, we hinted at the fact that our assumption that the first label set $L_1 = N$ is unnecessary.  While often convenient in practice, the assumption is arbitrary and obscures some mathematical properties of multilayer networks, as noted in \cite{Kivelae2015}.  We therefore propose a more general definition of both a multilayer network and its layers that lifts this constraint and give an example in which this definition is better suited to handle interesting queries of realistic data.

In the modified definition, we define a multilayer network as a tuple $\mathcal{M} = (V,E,w,l)$ with $l:V \to L_1 \times L_2 \times \cdots \times L_d$ a labeling function.  We do \emph{not} assume that $L_1$ is a distinguished set of nodes.  To form a layer of this network we choose $c\leq d$ many indices, ${j_1, j_2, \ldots, j_c}$ and we choose a vector of length $c$ given by $\alpha = (\alpha_{j_1}, \ldots , \alpha_{j_c})  \in L_{j_1} \times \cdots \times L_{j_c}.$  The vertices in layer $L_{\alpha}$ are $v\in V$ such that if $l(v) = (v_1, v_2, \ldots, v_d)$  then $v_{j_k} = \alpha_{j_k}$ for $k = 1,2, \ldots , c.$   That is, the vertices in layer $L_{\alpha}$ are those whose components in the $L_{j_k}$ are fixed and equal to the components of $\alpha$ and whose other components vary freely over the remaining labels.

\begin{figure}[ht!]
\begin{center}
\includegraphics{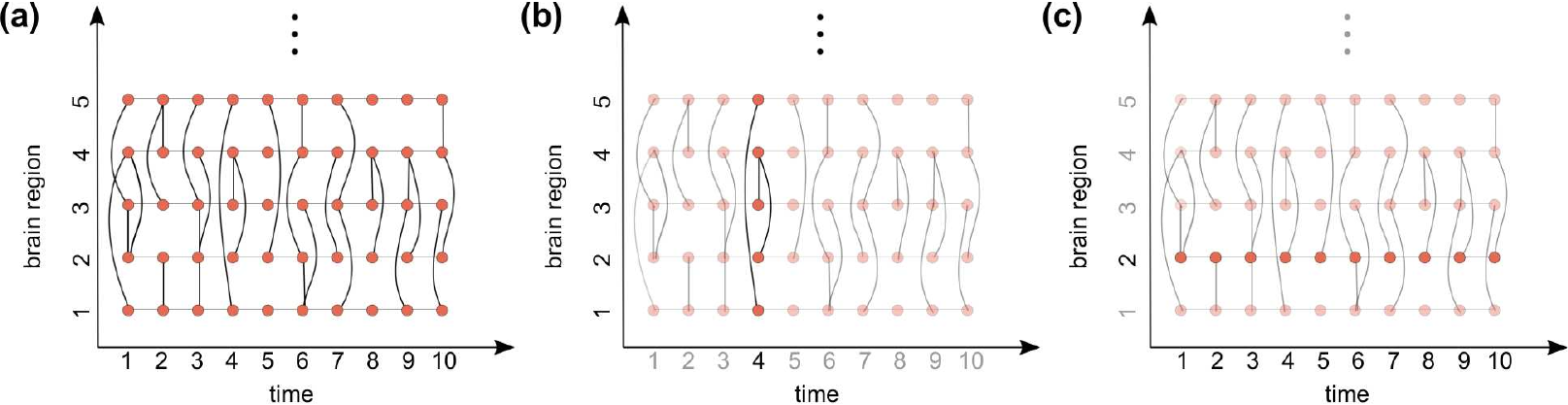}
\caption{\textbf{(a)}. Full temporal network. \textbf{(b)} Time layer at time 4. \textbf{(c)} Neuron Layer at neuron 2.}
\label{fig:temp_net_layers}
\end{center}
\end{figure}

Again consider the temporal network of section \ref{ex:temp_net2}.  If we fix a time $\alpha = (t_0)$ then we can form the time layer at time $t$ by taking all vertices $v$ such that $l(v) = (n, t_0).$  This is the usual partition of the network into layers based on time.  We could also fix a brain region $\alpha = (n_0)$ and form the $\alpha$ layer by taking all vertices $v = (n_0, t).$  This layer consists of a single brain region and all of its instances through time.  When the inter-layer edges are not measured empirically, but instead set as a constant $\omega$, this layer may not be particularly interesting.  Figure \ref{fig:temp_net_layers} shows examples of these two types of layers in panels (b-c).

\begin{figure}[ht!]
\begin{center}
\includegraphics{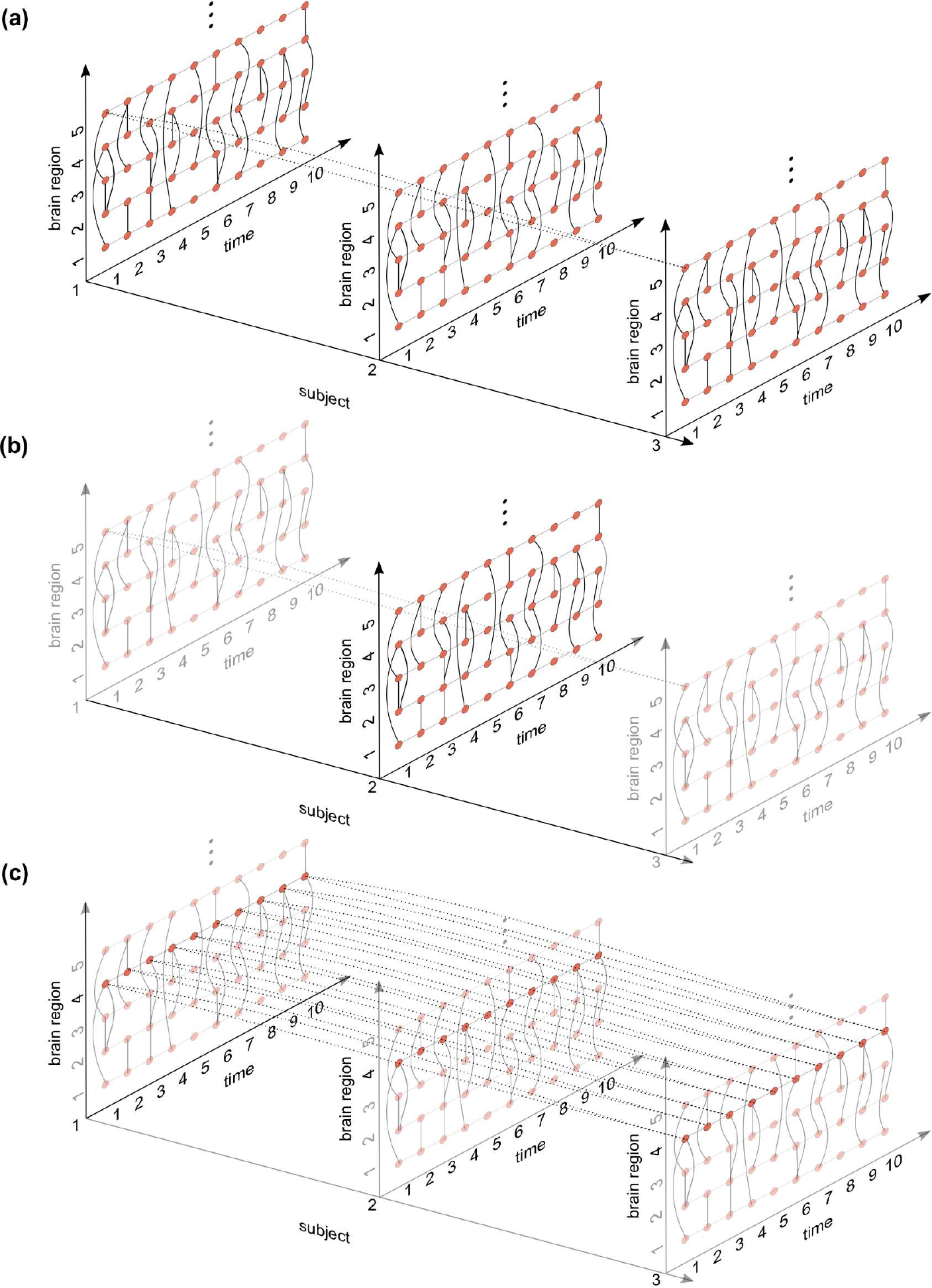}
\caption{\textbf{(a)}. Full 3 dimensional multilayer network.  For visual clarity edges between subjects are shown for only 1 brain region. \textbf{(b)} The layer associated with subject 2.  This layer is itself a multilayer network.  \textbf{(c)}  The layer associated with region 4.  Again this is a multilayer network. }
\label{fig:layers_subjects}
\end{center}
\end{figure}

Now consider a three dimensional multilayer network whose labels are brain regions, time, and subjects.  Suppose, like in the previous example, there are 50 brain regions, 10 time windows and assume there are 3 subjects.  We can choose many different layers to investigate.  For example, we can fix a subject, $\alpha = (s_0).$  The resulting layer is a temporal network corresponding to subject $s_0.$  Or we could fix both a subject and a time $\alpha = (t_0, s_0)$ and the resulting layer is the network of brain regions at time $t_0$ in subject $s_0.$  We could also fix a brain region and compare the activity across time and subjects.  See Figure \ref{fig:layers_subjects} for examples of these types or layers.

With this definition the layers of a multilayer network are linear slices of the dimensions of the network. This allows for some conceptual freedom in choosing and analyzing layers of the same multilayer network.  For example, when inter-layer edges are measured, instead of set as a parameter, it makes sense to consider the network formed by fixing a brain region and properties of this network might contain valuable information.  It also allows for mathematical freedom when investigating the structure of multilayer networks. The mathematics should be agnostic to the choice of labels for the nodes as the underlying structure of the network is agnostic to this choice.  We therefore encourage the neuroscience and general networks community to adopt this modified notation.

\section{Conclusion}
The human brain is a complex system of interacting units operating on multiple spatial and temporal scales with multiple modes of interaction.  As recording technology becomes more sensitive, diverse, and ubiquitous, and data sets become larger and richer so too must the method of analysis.  Multilayer networks are a simple, yet flexible data structure that provide unique quantitative analysis of complex data and have proven to be an indispensable tool to the neuroscientist due to their ability to retain information and structure otherwise lost using traditional approaches.  The retention of additional information combined with the development of novel network statistics has successfully unveiled insight into the structure and function of the human brain that was previously unobserved.  However, with these successes also come challenges.  Despite the utility of multilayer networks, to date, there are relatively few neuroscientific studies that incorporate the multilayer framework. It will be important for future research to utilize the ever expanding knowledge base and set of measures for multilayer networks as well as drive development of measures with improved sensitivity and specificity for the many potential applications.  The multilayer network framework has the potential to become the prominent mode of network analysis in the future, as neuroscientists face increasingly multi-modal, multi-temporal, or multi-scale data. Multilayer network science is in its infancy and comprehensive research into the structure and function of brain networks will be necessary as both multilayer networks and neuroscience develop in tandem.

\flushleft\textbf{Acknowledgements}
SFM would like to acknowledge support from the National Science Foundation (SMA-1734795) and the Army Research Laboratory (contract number W911NF-10-2-0022).  The content is solely the responsibility of the authors and does not necessarily represent the official views of any of the funding agencies.

\bibliographystyle{unsrt}
\bibliography{main}
\end{document}